\documentclass[12pt]{article}

\usepackage{amsmath,amssymb,amsfonts,amsthm,hyperref,bbm,latexsym,epsfig}
\usepackage{graphicx,multirow}
\usepackage{cite}
\usepackage{physics}

\newcommand{\bs}{\begin{subequations}}
\newcommand{\es}{\end{subequations}}
\newcommand{\be}{\begin{equation}}
\newcommand{\ee}{\end{equation}}
\newcommand{\ba}{\begin{eqnarray}}
\newcommand{\ea}{\end{eqnarray}}

\allowdisplaybreaks

\begin{document}

\title{\LARGE Unitarity Relations in the Presence of Vector-Like Quarks}

\author{\addtocounter{footnote}{2}
  Francisco Albergaria,\thanks{E-mail:
    {\tt francisco.albergaria@tecnico.ulisboa.pt}.}
  \ and\addtocounter{footnote}{1}
  G. C. Branco,\thanks{E-mail: {\tt gbranco@tecnico.ulisboa.pt}.}
  \\*[3mm]
  \small Centro de F{\'\i}sica Te\'orica de Part{\'\i}culas, CFTP, \\ 
  \small Departamento de F\'{\i}sica,\\ 
  \small Instituto Superior T\'ecnico, Universidade de Lisboa, \\
  \small Av.~Rovisco Pais~1, 1049-001 Lisboa, Portugal
  \\*[2mm]}

\date{\today}

\maketitle

\begin{abstract}
  We study, in a systematic way, the $V_{\text{CKM}}$ unitarity relations  which arise in extensions of the three generations Standard Model (3gSM) involving the addition of vector-like quarks (VLQ). In particular, we emphasize the effect of the presence of VLQ on $V_{\text{CKM}}$ moduli differences, as well on the size of the imaginary parts of rephasing invariant $V_{\text{CKM}}$ quartets. We consider the special case where an up-type VLQ is used to attempt at solving the unitarity problem in the first line of $V_{\text{CKM}}$.
\end{abstract}

\vspace*{4mm}

\section{Introduction}

In the Standard Model (SM), gauge invariance does not allow for the introduction of bare mass terms for quarks and leptons, since these terms are not gauge invariant. So quark and lepton masses only arise from Yukawa interactions, after spontaneous gauge symmetry breaking. The SM as introduced by Glashow, Salam and Weinberg predicts strictly massless neutrinos and therefore it has been ruled out by the discovery of neutrino oscillations \cite{Kajita2016,McDonald2016}, pointing towards non-vanishing neutrino masses. The simplest extension of the SM which can accommodate neutrino masses, consists of adding right handed neutrinos (RHN). Since RHN are $SU(2)$ singlets their Majorana mass can be much higher than the electroweak scale $v$. This leads immediately to the seesaw mechanism \cite{Minkowski1977,Gell-Mann1979,Yanagida1979,Glashow1980,Mohapatra1980,Mohapatra1981} and naturally small neutrino masses.

In the quark sector, one may have an analogous situation, one may also have quarks with bare mass terms which are gauge invariant. These quarks, usually called vector-like quarks (VLQ) (see \cite{Alves2023} for a review) are one of the simplest extensions of the SM. They may populate the desert between $v$ and the scale of grand-unification, without worsening the hierarchy problem. In the presence of VLQ the quark mixing matrix appearing in the charged currents is no longer an unitary matrix. For example, in the case of one $Q = -1/3$ vector-like quark, the quark mixing matrix consists of the first three lines of a unitary $4 \times 4$ matrix. 

In this paper we study unitary constraints which arise when VLQ are introduced in an extension of the SM. Unitarity constraints involving the standard quarks and VLQ are modified, since there is mixing between VLQ and standard quarks. We study in particular the effect of VLQ on CKM moduli differences. In the three generations SM one may define an asymmetry of the CKM matrix since all $|V_{jk}| - |V_{kj}|$ have the same value. This is no longer true when VLQ are present and we study the modifications which arise. In the 3 generations SM, the imaginary part of all rephasing invariant quartets have the same size which gives the strength of CP violation in the SM. This is no longer true when VLQ are
introduced and we study these effects.

The paper is organised as follows. In section \ref{sec:moduli}, we present some consequences of $3 \times 3$ unitarity, for $V_{\text{CKM}}$ moduli differences. We derive some relations for the moduli in the context of the SM and then show how these relations are modified in the presence of one up-type VLQ. In section \ref{sec:ims}, we analyse the implications of the presence of 1 VLQ of the up type for the size of the imaginary parts of rephasing invariant quartets. In section \ref{sec:pheno}, we recall how the so-called $V_{\text{CKM}}$ unitarity problem is solved through the addition of one up-type VLQ. In particular, we evaluate the imaginary parts of the rephasing invariant quartets showing that they are not all equal, as it is the case in 3-generations SM and we also evaluate the moduli differences and show that they can be different. Finally, in the last section, we present our conclusions.

\section{Moduli Differences}
\label{sec:moduli}

In the SM, the CKM matrix is a $3 \times 3$ unitary matrix. From normalization of the first row and first column of $V_{\text{CKM}} \equiv V$, one readily obtains
\be
|V_{12}|^2 - |V_{21}|^2 = |V_{31}|^2 - |V_{13}|^2.
\ee
Using the unitarity relations of the other rows and columns of $V$, we get
\bs \label{eq:2}
\allowdisplaybreaks
\begin{align}
   |V_{12}|^2 - |V_{21}|^2 &= |V_{23}|^2 - |V_{32}|^2,
   \\
   |V_{31}|^2 - |V_{13}|^2 &= |V_{23}|^2 - |V_{32}|^2.
\end{align}
\es
Within the three generations SM, one can introduce an asymmetry $\mathbf{a}$ defined as
\be \label{eq:3}
\mathbf{a} \equiv |V_{31}|^2 - |V_{13}|^2 = |V_{23}|^2 - |V_{32}|^2 = |V_{12}|^2 - |V_{21}|^2.
\ee
Experimentally, it is known that $\mathbf{a}$ is positive, since $|V_{31}| > |V_{13}|$.

For definiteness, consider an extension of the SM with one up-type VLQ. The quark mixing matrix is a $4 \times 3$ matrix which consists of the first three columns of a $4 \times 4$ unitary matrix
\be
V = \begin{pmatrix}
V_{11} & V_{12} & V_{13} & V_{14}
\\
V_{21} & V_{22} & V_{23} & V_{24}
\\
V_{31} & V_{32} & V_{33} & V_{34}
\\
V_{41} & V_{42} & V_{43} & V_{44}
\end{pmatrix}.
\ee
From unitarity of the first row and first columns of the $4 \times 4$ matrix, we have
\bs
\allowdisplaybreaks
\begin{align}
   |V_{11}|^2 + |V_{12}|^2 + |V_{13}|^2 + |V_{14}|^2 &= 1, \\
   |V_{11}|^2 + |V_{21}|^2 + |V_{31}|^2 + |V_{41}|^2 &= 1.
\end{align}
\es
Subtracting these equations, we get
\be
\label{eq:a1213}
a_{12,13} \equiv \left(|V_{12}|^2 - |V_{21}|^2 \right) - \left(|V_{31}|^2 - |V_{13}|^2 \right) = |V_{41}|^2 - |V_{14}|^2.
\ee
Applying a similar procedure using the unitarity relations of the other rows and columns of $V$, we get
\bs
\label{eq:a1232}
\allowdisplaybreaks
\begin{align}
   a_{12,32} \equiv \left(|V_{12}|^2 - |V_{21}|^2 \right) - \left( |V_{23}|^2 - |V_{32}|^2 \right) &= |V_{24}|^2 - |V_{42}|^2,
   \\
   a_{13,23} \equiv \left(|V_{13}|^2 - |V_{31}|^2 \right) - \left( |V_{32}|^2 - |V_{23}|^2 \right) &= |V_{34}|^2 - |V_{43}|^2.
\end{align}
\es

From $D_0 - \overline{D_0}$ mixing, we know that, in models with one up-type VLQ, we have \cite{Alves2023}
\be
\label{eq:d0mixing}
|V_{14}|^2 |V_{24}|^2 < \left(2.1 \pm 1.2\right) \times 10^{-8}.
\ee

\section{Differences between the imaginary parts of the quartets}
\label{sec:ims}

In the SM, if we use unitarity between the first two rows of the CKM matrix, we get
\be
V_{11} V_{21}^* + V_{12} V_{22}^* + V_{13} V_{23}^* = 0.
\ee
Multiplying this equation by $V_{11}^* V_{21}$, we get
\be
|V_{11}|^2 |V_{21}|^2 + V_{12} V_{21} V_{11}^* V_{22}^* + V_{13} V_{21} V_{11}^* V_{23}^* = 0.
\ee
Taking the imaginary part of this equation we get
\be
\Im Q_{1221} = \Im Q_{1123},
\ee
where $Q_{\alpha i \beta j} \equiv V_{\alpha i} V_{\beta j} V_{\alpha j}^* V_{\beta i}^*$. We can apply a similar procedure with different rows or columns of the CKM matrix and we obtain the well known result that the imaginary parts of all the quartets in the CKM matrix are equal up to a sign.

Consider now the case of a model with one up-type VLQ. We have
\be
\sum_{i=1}^4 V_{\alpha i} V_{\beta i}^* = \delta_{\alpha \beta}.
\ee
Consider the case where $\alpha \neq \beta$. Multiplying the equation above by $V_{\alpha j}^* V_{\beta j}$, we get
\be
\sum_{i=1}^4 V_{\alpha i} V_{\beta i}^* V_{\alpha j}^* V_{\beta j} = 0.
\ee
Taking the imaginary part of this equation we get
\be
\sum_{i \in \{1,2,3,4\} \setminus j} \Im Q_{\alpha i \beta j} = 0.
\ee
Suppose that $j \neq 4$. We get
\be
\sum_{i \in \{1,2,3\} \setminus j} \Im Q_{\alpha i \beta j} - \Im Q_{\alpha j \beta 4} = 0.
\ee
Defining $F_{\alpha \beta} \equiv \delta_{\alpha \beta} - V_{\alpha 4} V_{\beta 4}^*$ as the FCNC matrix, we can rewrite the equation above as
\be
\label{eq:QF}
\sum_{i \in \{1,2,3\} \setminus j} \Im Q_{\alpha i \beta j} = \Im \left(V_{\beta j} V_{\alpha j}^* F_{\alpha \beta}\right).
\ee
If $\alpha \neq 4$ and $\beta \neq 4$, then on the left-hand side of this equation we will have two quartets involving elements of the top-left $3 \times 3$ block of the CKM matrix and we can see that their sum, which in the SM is equal to $0$, will depend on the FCNC matrix elements. Putting these to zero (as is the case in the SM), we recover the fact that the moduli of the imaginary part of these quartets involving elements of the top-left $3 \times 3$ block of the CKM matrix become equal.

If, for example, we choose $\alpha = 1$, $\beta = 2$ and $j = 1$, we get
\be
\label{eq:im1}
\Im Q_{1221} - \Im Q_{1123} = \Im \left(V_{21} V_{11}^* F_{12} \right).
\ee

Similarly to how we obtained Eq. \eqref{eq:QF}, we can get
\be
\sum_{\alpha \in \{1,2,3\} \backslash \beta} \Im Q_{\alpha i \beta j} = \Im Q_{\beta i 4 j}.
\ee
If, for example, we choose $\beta = 1$, $i=1$ and $j = 2$, we get
\be
\label{eq:im2}
\Im Q_{2112} - \Im Q_{1132} = \Im Q_{1142}.
\ee

\section{Phenomenology}
\label{sec:pheno}

\subsection{CKM Unitarity Problem}


Recently, using improved values for the form factors and radiative corrections associated to the relevant meson and neutron decay processes, a deviation from the unitarity condition of the first row of the CKM matrix was found. This discrepancy is known as CKM unitarity problem \cite{Seng2018,Seng2019,Czarnecki2019,Belfatto2020,Seng2020,Cheung2020,Branco2021,Hayen2021,Shiells2021,Belfatto2021,FLAG2022,Balaji2022,Accomando2022,Crivellin2022,Belfatto2023} and is one of the main motivations for the introduction of VLQ extensions of the SM. The experimental results are compatible with
\be
|V_{ud}|^2 + |V_{us}|^2 + |V_{ub}|^2 = 1 - \epsilon,
\ee
where \cite{Alves2023}
\be
\sqrt{\epsilon} = 0.04 \pm 0.01 \ (95 \% \ \text{C.L.}).
\ee

The introduction of an up-type VLQ can solve the CKM unitarity problem as the sum of the squares of the moduli of the elements of the first row of the CKM matrix is, in this case, equal to $1 - |V_{14}|^2$. Thus, $|V_{14}|^2$ may account for $\epsilon$.

Consider the following parametrization of a $4 \times 4$ unitary matrix, of which, in the case of a model with one up-type VLQ, the CKM matrix are the first three columns \cite{Botella1986}:
\be
V = \left(\begin{array}{@{}c c@{}}
  V_{\text{PDG}} & 
  \begin{matrix}
  0 \\
  0 \\
  0
  \end{matrix} \\
  \begin{matrix}
  0 & 0 & 0
  \end{matrix}
  & 1
\end{array}\right) \begin{pmatrix}
    c_{14} & & & - s_{14} e^{- i \delta_{14}}
    \\
     & 1 & & 
    \\
      & & 1 & 
    \\
    s_{14} e^{i \delta_{14}} & & & c_{14}
\end{pmatrix} \begin{pmatrix}
    1 & & &
    \\
     & c_{24} & & - s_{24} e^{- i \delta_{24}}
    \\
      & & 1 & 
    \\
     & s_{24} e^{i \delta_{24}} & & c_{24}
\end{pmatrix} \begin{pmatrix}
    1 & & &
    \\
     & 1 & &
    \\
     &  & c_{34} & - s_{34}
    \\
     &  & s_{34} & c_{34}
\end{pmatrix},
\ee
where $V_{\text{PDG}}$ is the standard PDG parametrization of a $3 \times 3$ unitary matrix \cite{pdg2022},
\be
V_{\text{PDG}} = \begin{pmatrix}
    c_{12} c_{13} &  s_{12} c_{13} & s_{13} e^{-i \delta} \\
    -s_{12} c_{23} - c_{12} s_{23} s_{13} e^{i \delta} & c_{12} c_{23} - s_{12} s_{23} s_{13} e^{i \delta} & s_{23} c_{13}
    \\
    s_{12} s_{23} - c_{12} c_{23} s_{13} e^{i \delta} & - c_{12} s_{23} - s_{12} c_{23} s_{13} e^{i \delta} & c_{23} c_{13}
\end{pmatrix}.
\ee

In the framework of solving the CKM unitarity problem, we consider $s_{14} = 0.04$. From Eq. \eqref{eq:d0mixing}, we find then $s_{24} \lesssim 3.63 \times 10^{-3}$. We consider the PDG values \cite{pdg2022}
\bs
\allowdisplaybreaks
\begin{align}
   s_{12} &= 0.22500 \pm 0.00067,
   \\
   s_{13} &= 0.00369 \pm 0.00011,
   \\
   s_{23} &= 0.04182^{+0.00085}_{-0.00074},
   \\
   \delta &= 1.144 \pm 0.027.
\end{align}
\es
Assuming $\frac{1}{5} s_{24} \lesssim s_{34} \lesssim 5 s_{24}$, we find for the moduli differences
\bs
\begin{align}
    - 10^{-4} \lesssim & \, a_{12,13} \lesssim - 10^{-6},
    \\
    - 10^{-4} \lesssim & \, a_{12,32} \lesssim - 10^{-5},
    \\
    - 10^{-5} \lesssim & \, a_{13,23} \lesssim 10^{-5},
\end{align}
\es
where $a_{12,13}$ was defined in Eq. \eqref{eq:a1213} and $a_{12,32}$ and $a_{13,23}$ were defined in Eq. \eqref{eq:a1232}.

Making the same assumption for $s_{34}$ and using Eqs. \eqref{eq:im1} and \eqref{eq:im2}, we get for the differences between the imaginary parts of the quartets
\bs
\begin{align}
   &|\Im Q_{1221} - \Im Q_{1123}| \lesssim 3 \times 10^{-5},
    \\
    & |\Im Q_{2112} - \Im Q_{1132}| \lesssim 4 \times 10^{-5}.
\end{align}
\es

\section{Conclusions} 

We study, in a systematic way, how the presence of Vector Like Quarks (VLQ) changes the unitarity relations which hold in the three generations SM (3gSM). For example, it is well known that in the 3gSM, the moduli of the imaginary parts of all $V_{\text{CKM}}$ rephasing invariant quartets have the same size. This is no longer true in extensions of the SM with VLQ. These imaginary parts of rephasing quartets no longer have the same size and the variations in the size are related to the appearance of FCNC. In the case of an up-type VLQ, the most relevant FCNC are those involving the $cu$ FCNC which are constrained by $D_0$-$\overline{D_0}$ mixing and the $ct$, or $ut$  currents which are constrained by bounds on rare top quark decays. We analyse how the relations between the imaginary parts of the quartets change in the case of an extension with one up-type VLQ.

In the 3gSM one may define an asymmetry denoted $\mathbf{a}$ which is related to the fact that all the differences of squared $V_{\text{CKM}}$ moduli have the same value (see Eqs \eqref{eq:2} and \eqref{eq:3}). This is no longer true in VLQ extensions of the 3gSM. We also analyse how these moduli differences change in the case of an extension with one up-type VLQ.

It is well known that an improvement on the precision of experimental results, leads to a deviation of unitarty in the first row of the CKM matrix. It has been pointed out that this problem can be solved through the addition of an up-type VLQ. We analyse, in the conditions necessary to solve this problem, the possible variations in the size of imaginary parts of rephasing invariant $V_{\text{CKM}}$ quartets, as well as the possible variations in the differences of the moduli of the CKM matrix elements.

\section*{Acknowledgments}
The authors thank the Portuguese Foundation for Science and Technology (FCT)
for support through the projects UIDB/00777/2020,
UIDP/00777/2020,
and CERN/FIS-PAR/0002/2021.
The work of F.A.\ was furthermore supported by grant UI/BD/153763/2022.

\newpage




\end{document}